# Avoiding the "Great Filter": Extraterrestrial Life and Humanity's Future in the Universe

Jonathan H. Jiang[1], Philip E. Rosen[2], Kelly Lu[3], Kristen A. Fahy[1], Piotr Obacz[4]

[1]Jet Propulsion Laboratory, California Institute of Technology, Pasadena, California, USA
[2]Independent Researcher, Vancouver, WA, USA
[3]Santa Margarita Catholic High School, Rancho Santa Margarita, CA, USA
[4]Faculty of International and Political Studies, Jagiellonian University, Krakow, Poland

Correspondence: Jonathan.H.Jiang@jpl.nasa.gov

Keywords: extraterrestrial, life, civilization, humanity, filter, future

## Abstract

Our Universe is a vast, tantalizing enigma - a mystery that has aroused humankind's innate curiosity for eons. Begging questions on alien lifeforms have been thus far unfruitful, even with the bounding advancements we have embarked upon in recent years. Coupled with logical assumption and calculations such as those made by Dr. Frank Drake starting in the early 1960s, evidence of life should exist in abundance in our galaxy alone, and yet in practice we've produced no clear affirmation of anything beyond our own planet. So, *where is everybody?* The silence of the universe beyond Earth reveals a pattern of both human limitation and steadfast curiosity. Even as ambitious programs such as SETI aim to solve the technological challenges, the results have thus far turned up empty for any signs of life in the galaxy. We postulate that an existential disaster may lay in wait as our society advances exponentially towards space exploration, acting as the Great Filter: a phenomenon that wipes out civilizations before they can encounter each other, which may explain the cosmic silence. In this article, we propose several possible scenarios, including anthropogenic and natural hazards, both of which can be prevented with reforms in individual, institutional and intrinsic behaviors. We also take into account multiple calamity candidates: nuclear warfare, pathogens and pandemics, artificial intelligence, meteorite impacts, and climate change. Each of these categories have various influences but lack critical adjustment to accommodate to their high risk. The Great Filter has the potential to eradicate life as we know it, especially as our rate of progress correlates directly to the severity of our fall. This indicates a necessary period of introspection, followed by appropriate refinements to properly approach our predicament, and addressing the challenges and methods in which we may be able to mitigate risk to mankind and the nearly 9 million other species on Earth.

## 1. Introduction

Life, in as much as has been determined from extensive sampling of only a single world, poses a dilemma. In a universe whose normal matter is almost entirely hydrogen – with just a single proton, the simplest of the elements - humanity finds within itself and its environment a wealth of chemical complexity which seems to defy logic. The solution to this riddle of higher development is found, oddly enough, in grand scale destruction. In their explosive demise, stellar furnaces fuse together heavy nuclei which, upon combining with electrons, enable exponentially branching combinatorial chemistry including the formation of large molecules. In this realm hydrogen is rendered merely a bit player, ubiquitous but no longer occupying center stage. Rather, the six-proton nucleus of carbon, with its particular arrangement of electrons given over to orbital hybridization, seizes the central role in biochemistry.

But, if "life as we know it is merely an afterthought in the global scheme of the cosmos" [1], we may well be led to conclude the Earth's bounty is of truly extraordinary – perhaps even unique – nature. Are humans, just one among the millions of species sharing this remarkably hospitable but fragile oasis in the cosmos, a kind of multiply improbable instance of fused ash first assembling to primitive life then, much later, tumbling into self-awareness followed by spacefaring technological prowess? One school of thought posits this 'Rare Earth' hypothesis [2]: given a Universe stretching approximately 92 billion lightyears and existing for nearly 14 billion years, intelligent life can be both inevitable but still exceedingly rare. Hence, present era Earth is merely the particular time and place such extraordinarily long odds, in effect, paid off and we are the lucky beneficiaries. While such a notion may come as comfort to some as they (philosophically speaking) claim universal ownership, this scenario would also leave us profoundly isolated and stunted. The great scientists, mathematicians and artists our civilization has produced achieved their historic feats through collaboration and competition. Extending this notion beyond our home world, how could humanity as a species ever truly realize our full potential if there are no other technological civilizations with whom to interact?

Technological developments in the years following Enrico Fermi's famous question, posed to colleagues in 1950 and forming the Paradox [3] which came to bear the name of the great 20$^{th}$ century physicist, enabled the search for extraterrestrial intelligence (SETI) to commence. Among these innovations are radio astronomy and dramatic developments in rocketry and computational power. Astronomers, long confined to making observations within just the very narrow band of the electromagnetic spectrum afforded by human vision, could now view and measure a cosmos whose radiation signatures extended from long wave radio through high energy x-rays. For centuries the light reaching Earthly observers from the planets and other bodies of our Solar System was no more than tiny atmosphere-distorted points and weakly resolved discs flickering in glass lenses. Many of these worlds and other celestial objects have now been physically visited by humanity's robotic emissaries while Earth's Moon has been touched by humans themselves. Additionally, our vantage point in observing the Universe has taken up positions off our home world, using ever more sophisticated and sensitive instruments to peer across billions of lightyears and even in a few cases imaging worlds orbiting other suns. Since 1992 over 5000 exoplanets have been confirmed with several thousand candidates additionally pending, attesting to the ubiquitous nature of planetary systems. Moreover, modeling from such early works as the Drake Equation [4] to more recent investigations suggest extraterrestrial intelligence may well have arisen in the Milky Way [5], [6], [7]. Pursuing the question still further, in recent years serious exploration of the complex implications for human society upon coming into contact with life off the Earth has moved into the realm of mainstream scientific inquiry [8]. If resolution of the Fermi Paradox does not to condemn humanity to a lonely Universe, it is only logical that life must inevitably strive to seek other life wherever it may exist. Or as Carl Sagan more eloquently proffered, "In the deepest sense the search for extraterrestrial intelligence is a search for ourselves."

Just as technology enables humanity to push back the boundaries of our knowledge of the cosmos, it tempts as well with the means of self-destruction. Abruptly realized, we find ourselves the sole stewards of a resource rich world – and socially ill-prepared for the job. A worrying sense of humanity's technological cleverness outpacing our better judgment pervades, for even as we reach far beyond Earth's gravity well we are being pulled down by internal strife. This presents a potentially universal question, as well as a most unsettling solution to Fermi's Paradox. As depicted in Figure 1, is there a "Great Filter" [9] of sorts awaiting every civilization which sets out on the path of technological development and if so, has humanity yet to confront this ultimate rite of passage? An optimistic perspective would point out our continued existence despite first developing the capability for self-annihilation in 1945. Caution, however, is well warranted as these past 77 years (in the developed world, merely an average human lifespan) have been fraught with near misses such as the Cuban Missile Crisis in 1962 and persistent flaring of armed conflicts around the globe. Additionally, human activity has unsettled the Earth's otherwise highly accommodative environment for supporting life, casting a dark shadow over the prospect of endlessly advancing technological innovation opening up an unlimited future for humanity to spread across space and time. Returning to the Drake Equation, recent modeling suggests that it is the lifespan of civilizations capable of interstellar communication, "L", which is the most influential among its seven variables [5], [10]. Taking this claim as stipulation, it follows immediately that the sub-factors comprising L must be identified and studied in detail if we are to maximize humanity's lifespan.

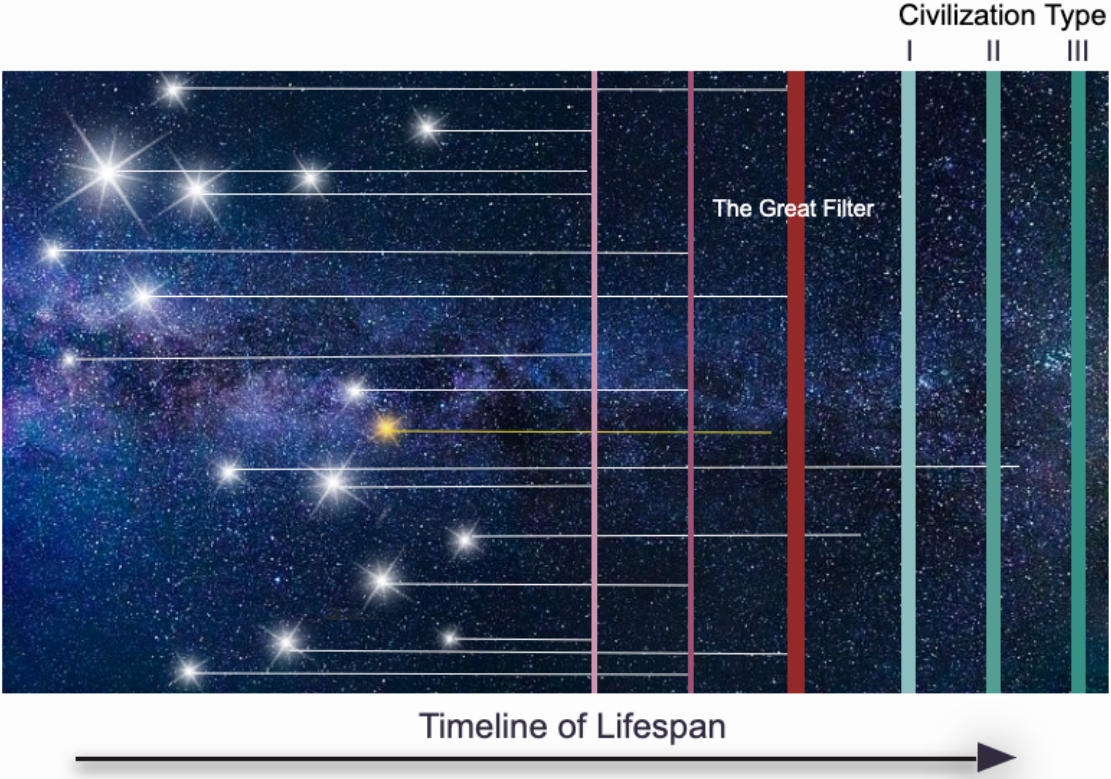

**Figure 1:** Depicts an example timeline of potential lifespans for intelligent life in the galaxy. There exists no pattern for when life may arise, though we postulate that various "filters" as shown in the vertical pink lines may have caused mass extinction of other civilizations. Our Sun and its subsequent line, shown in yellow, is a representation of the amount of time humankind has been able to survive up until present. Life on Earth has already overcome filtering events, however the Great Filter lies ahead and it is unknown if we will be able to survive long enough to become a Type I civilization or beyond (green vertical lines), as others already may have or have not.

In pragmatically simplified terms, the notion of a filter suggests a structure specifically configured to allow one or more constituents of a multiple component flow to pass through while the remaining constituents are prevented from fully crossing to the far side. In this analogy, the overall development (i.e., forward-moving flow) of species' evolution to sentience, as well as the follow-on development of civilization itself, is the bulk fluid and the critical characteristics particular to each component species determines if it passes this semi-permeable barrier or is captured, irreversibly halting progress. The key to humanity successfully traversing such a universal filter is found in understanding what characteristics the barrier will constrain, identifying those attributes in ourselves and neutralizing them in advance. Human civilization over the past 5000+ years, and in particular since 1945, has revealed much of what would surlily impede, if not outright arrest, our aspirations to colonize other worlds in the Solar System and beyond. It seems as though nearly every great discovery or invention, while pushing back the borders of our technological ignorance, is all too quickly and easily turned to destructive ends. Examples such as splitting the atom, biomedical innovations and resource extraction and consumption come to mind with disconcerting swiftness. Still, some have suggested artificial intelligence (AI) as yet another factor which, pending substantial technical hurdles, may yet have its chance to prove friend or foe. The implications of AI pose complexities perhaps only an AI itself could understand, and has even been suggested as an alternative explanation [11] for Fermi's Paradox. Although alien AI is, inescapably, a doubly strange notion, some perspective may be gained when compared to still more exotic hypotheses such as involving dark matter [12]. Finally, nature itself has the potential to extinguish human civilization, often envisioned as an asteroid or comet impact similar to that which Earth suffered 66 million years ago, triggering the Cretaceous-Paleogene extinction and wiping out roughly 75% of all species worldwide.

## 2.0 Rationale

If life arisen on Earth is ever to know of life elsewhere, assuming such exists, we as the Earth's sole technological species must first come fully to terms with ourselves and our environment. The struggle for survival, security and dominance - all rooted in human passions - drives creativity and with it, civilization and invention [13]. As history has shown time and again, however, this cleverness comes at great cost. The human brain, still orders of magnitude more complex in terms of synaptic connections than the transistor-based structures underlying the most advanced supercomputers, holds the key. Using our demonstrated inventiveness to proactively recognize, diagnose and formulate countermeasures to the most serious threats to our existence, humanity may yet avert the Great Filter. In so doing we would likely emerge downstream of the Great Filter as a near Type I civilization on the Kardashev scale [14], ready to seek our place in a future greater than what we could realize if confined to just our home world. Indeed, recent modeling suggests human-crewed exploration of our Solar System beyond Mars may well be possible within this century [15], [16]. Analysis of these leading threats has found them to include large-scale nuclear warfare, pathogens (both naturally occurring and engineered), artificial intelligence, impacts from asteroids and climate change [17].

Unchecked population growth is a factor in the aforementioned scenarios, excepting that of asteroid impact, threatening human civilization and life on Earth in general. At present world population is nearing 8 billion, an exponential rise from about 1.6 billion at the start of the 20$^{th}$ century, and has doubled over the last 49 years. While Malthusian-inspired worst-case predictions of such a rapid increase [Ehrlich, 1968] have so far been averted, thanks in large part to technological advancements in farming, energy production and distribution, invention cannot be expected to indefinitely offset the multifaceted stresses imposed by ever escalating population. However, further improvements in modeling and better-informed controls, with education in developing nations as a critical factor for success, suggest a pathway to towards reducing population at a modest pace after a projected peak of slightly less than 10 billion is passed in the 2060s [18]. If such a prediction is at least directionally correct, it is not unreasonable to cautiously expect a moderating effect will ensue in many of the major challenges humanity now faces as we move towards the latter decades of the 21st century.

2.1 Nuclear War

Warfare has beset humanity long before civilization began taking hold approximately 5200 years ago. An outgrowth of tribalism and our innate sense of competition for resources, invention has been harnessed to make successively more deadly weapons, as well as defenses. With the passage of time, the horizon has effectively drawn closer to home and no more geographical frontiers remain to be to conquered here on our native world. War cannot be thought of as "something bad" happening to unfortunate people a safe distance away from where one has chosen to live. That said, there are encouraging signs of rationality emerging: peace agreements in the historically troubled Middle East, a vast reduction in nuclear warheads since the height of the Cold War and a wide coalition of nations rallying their support for the besieged in Eastern Europe. Amid continued strife it should also be noted that at the start of the 20th century the U.S. was among a very small handful of constitutional democracies in the world. The present finds a clear majority of the world's nations are, albeit imperfectly and often only partially, democracies with at least nominally representative governments [19]. Though strained, the Democratic Peace Theory - which holds that democracies are hesitant to go to war with each other [20] - has historically been borne out. In parallel to this hopeful trend, the end of WWII saw the age of colonial rule finally begin its long ebb into tragic history. Hence, to estimate the threat of large-scale warfare one may take as proxy the quantified extent of constitutional democracies across the world with time. Weighting relatively more heavily the trends towards or away from functionally representative democracy in those nations controlling significant stockpiles of nuclear weapons would logically follow. The past, it has often been said, is prologue - but it need not be prediction. A future free of catastrophic warfare remains, for now, within humanity's grasp and with it, avoidance of perhaps the most obvious of Great Filters.

2.2 Pathogens & Pandemics

As very recent events have painfully reminded us, biologically-based threats remain at the forefront of humanity's many concerns. Pathogens, microscopic but with potential for causing death on a planetary scale, have continually emerged throughout history. Although the vast majority of viruses and bacteria are either harmless or nearly so, combinatorial biochemistry incubated in large populations and integrated across a great many iterations have nonetheless repeatedly given rise to the rare deadly strain capable of rapid transmission. Catalyzing this threat in modern times is human civilization's ever-increasing interconnectedness, shrinking the vast distances between continents to effectively that which might have existed between neighboring medieval villages. It is not unreasonable to suggest humanity has actually been fortunate to have only encountered two particularly serious pandemics since WWI. Here again, our ingenuity holds the means of survival. Whereas past generations were at the mercy of deadly pathogens, modern diagnostic and pharmaceutical techniques are powerful allies in containing and, ultimately, defeating this recurring foe. The current struggle with SARS-CoV-2 offers a silver lining: an opportunity to model quantitatively the factors which comprise the threat of pandemic in the setting of a technologically well-equipped society. Data from past pandemics such as Influenza, which struck in the years immediately after WWI, may well add some useful (albeit limited) context to current modeling of pandemics. That said, it must be emphasized there were no vaccines quickly developed to battle that pandemic and it ran its horrific course unencumbered just as pandemics had done long before the turn of the 20th century. In summary, application of reliable data *in the present* [21] must take centerstage in predicting how future pandemics will spread, how deadly they will be and how quickly and effectively we will be able to leverage our knowledge of the life sciences to counter this manifestation of the Great Filter.

2.3 Artificial Intelligence

Once confined to the realm of speculative fiction in popular works such as Arthur C. Clarke's *2001: A Space Odyssey* (1968), William Gibson's *Necromancer* (1984) and James Cameron's *The Terminator* (1984), the practicality of achieving artificial intelligence has moved methodically towards realization with advances in microcircuit technology. While dramatic screen and print depictions range from that of friendly androids to malevolent supervillains bent on world destruction, a sober assessment of the actual risks posed by AI remains as elusive to full comprehension as the minds of AI's presumptive human inventors. Taking the guarded view, if and when AI does come to fruition it may well be too late to rely on empirical evidence gathered from its actual attitudes and behaviors towards the species which brought about its existence. Prudence then strongly suggests we perform sooner rather than later what modeling can be done, evaluate the necessarily preliminary conclusions drawn and proactively plan for a peaceful approach to the possibility of sharing the Earth with a new technological entity. To start, an assumption is required which theorizes arrival of AI is conditionalized, though not guaranteed, on achieving with hardware the same level of structural complexity as that of the human brain, which itself encompasses $\sim 10^{14}$ synaptic connections among its $\sim 10^{11}$ neurons [22]. Presently, microprocessors can contain up to $\sim 10^{10}$ transistors - electronic gateways which serve a roughly comparable function to the bioelectrically driven synaptic connections between neurons in the brain [23], [24]. As the density of transistors per microprocessor has increased exponentially since the 1960s, this corresponding to what is widely characterized as Moore's Law [25], one can project when computer sophistication may rival that of the human mind. Of course, impediments posed by material limitations and quantum effects would need to be overcome if this rapid, decades-long trend towards brain-like complexity is to be maintained beyond the mid-2020s. As for whether AI would be benign or otherwise, self-imposing a Great Filter of our own invention, that will depend on the evolving nature and disposition of Earth's first high-tech species.

2.4 Asteroid & Comet Impacts

For centuries astronomers considered the movements of the Solar System's visible planets and the stars to mimic that of clockwork. Indeed, primitive timekeeping depended on the comings and goings of the Sun, Moon, "wandering stars" (i.e., planets) and those flickering specs of light - many of which we now know warm worlds many lightyears distant. The lesser remains of our Solar System's formation still orbit the Sun as asteroids in their uncounted billions, carbonaceous or stony objects while still others contain significant percentages of metals. Gravitational perturbations sometimes send these remnants sunward, typically originating from outer regions such as the Kuiper Belt and the Oort Clouds, where they can occasionally tumble across planetary orbits. Analogously, the paths of comets – icy bodies sheathed in frozen gases – may very occasionally intersect planets and moons. Most objects are relatively small and upon encountering Earth's atmosphere at high velocity, disintegrate into harmless bits or cinders. There are, however, a non-zero percentage which are large enough to survive passage through the atmosphere and, impacting the surface, cause catastrophic destruction to our sensitive biosphere. Risk modeling for such ultra-low frequency, high severity events are, by their very nature, challenging. At the fundamental level such calculations involve the product of especially low annualized likelihoods multiplying difficult-to-fathom severity factors to produce time dependent cumulative risk trends. A useful information source which may be leveraged for this purpose is found in one of the major initiatives addressing this challenge to planetary defense: NASA's Near-Earth Object (NEO) Observations Program [26]. A simplified procedure would envision mining public-facing data from this resource and making use of readily available risk management modeling techniques to generate quantifiable results. The end product would be weighted risk curves versus time for mass extinction event (MEE) level impacts – i.e., a vanishingly tiny risk of a MEE in the next year ranging asymptotically towards 100% as likelihood is integrated over time into the very distant future. It should be noted that as with any statistically derived result there is no guaranteed interval of complete safety, only a likely window to prepare in the near term. Fortunately, NEO and other projects such as the Double Asteroid Redirection Test (DART) mission are examples of humanity using our technological capabilities to proactively address this possible version of the Great Filter.

2.5 Climate Change

In recent decades there is perhaps no large-scale threat to life on Earth which has been studied more intensely than climate change. While public opinion on the bottom-line implications of a warming biosphere continues to vary, general acceptance of the basic contention that surface temperatures are rising and human activity is a significant driver has largely moved beyond doubt among national governments. Central to this focus is the United Nations' Intergovernmental Panel on Climate Change (IPCC), whose ongoing investigations include updated predictions for rising temperatures resulting from emissions of greenhouse gases (GHGs) using a multitude of climate models [27]. Many independent studies have also been performed such as those seeking an empirically-based relationship between surface temperature rise and escalating concentration of the chief GHG, $CO_2$, in the lower atmosphere [28]. Given this plethora of models, logic strongly suggests the most reliable method for making analytically sound predictions for climate change is not in piecing together new models but rather determining where the most widely accepted established models converge – i.e., a preponderance of the evidence approach. The relatively slow nature of climate change, along with disagreements between some of the models' predictions, continues to present headwinds to engaging wider efforts to blunt the effects of GHGs. The major impediment to taking more decisive actions, however, are the challenges imposed by transitioning to non-carbon-based energy sources such as solar, wind, nuclear power. Here again, rapidly advancing technologies in areas such as modularized nuclear power plants [29] and carbon capture and sequestration (CCS) are among the best hopes for avoiding slow-motion ensnarement by this lulling but lethal Great Filter.

**3.0 Conclusions and Discussion**

"In the deepest sense, the search for extraterrestrial intelligence is a search for ourselves." A succinct quote expressed by Carl Sagan over forty years prior still holds veritable truth in our modern age. It's impossible to deny that one of the most tantalizing mysteries of our worldly existence lies in what exists beyond it—specifically, the potential of uncovering alien life. Many great minds and the more ordinary alike have approached this subject from various angles. Some take avenues of media, others leverage avant-garde technology as a means of understanding, while still others fill the spectrum in between with assumptions of their own construction. To the current date, no substantial or particularly promising traces of intelligent life have been detected. This (apparent) absenteeism makes the idea of "aliens" all the more tantalizing.

But perhaps the persistence of effort towards the field reveals underlying sides to our own motives. The idea of being alone in a universe vaster than our creativity can touch is terrifying to fathom: a feeling of cosmic isolation. And the postulation of a phenotypically unique organism having the intelligence to communicate, or at least leaving evidence of substance, is fascinating. If an octopus opening a jar or an elephant brushing some paint strokes is enough to catch the eye of billions, discovery of sentience beyond our biosphere would send global shockwaves.

Thinking big picture, the prospective discovery of extraterrestrial intelligence can be viewed philosophically with roots in theology. *Why are we so perseverant?* Often, it's a matter of quasi-Darwinian instinct and survival of the fittest, if "fit" were to be a shared standard and "instinct" a scope for rivalry. If indeed aliens don't exist and perhaps never had, we would find ourselves as the sole ringleader of the foreseeable universe with no outsider competition. Humanity, speaking secularly, would have been an accident from the start, a fluke of animated chemistry called "life" in an otherwise vast emptiness. Thusly, our freedom to explore would be confined only by our own shortcomings.

Some have argued against efforts of exploration. Besides lingering feelings of curiosity, hope, and even imperialistic desire, that childish wonderment is buried under the avalanche of our inadequacies. Playing galactic leader will only come to fruition if we can proposition and bring to execution a united front. To broaden a famous saying, this for the task ahead, "only together can we hope to stand against the cosmos but divided we fall quickly back to Earth, never to know others". There is no guarantee what kind of relationship would develop with beings so cryptic and distant from ourselves. First contact may very well abolish our fragmented society, especially the scattered and fragile coalition in which humans have organized themselves.

This disunity and admitted dysfunction may snowball quickly into the Great Filter. If, unfortunately, this filter exists ahead of us we have some sizeable challenges to overcome, even as the catalyst of this astral pattern may actually lay within ourselves. The foundation for many of our possible filters find its roots in immaturity. Warfare in itself is absurd - the loss of human life for the benefit of few. Governments pour money into weaponry that costs billions, targeting the erasure of the priceless thing of human life. This type of fundable propaganda has led to the militarization of society, perpetuating fatal conflict as a solution to human-made problems, erecting cultural barriers which perpetuate an infamously vicious circle.

This immaturity is present in other Great Filter candidates. Poor distribution of resources and slow responses to inefficient regulations have led to profound shortcomings in the solutions to our recent pandemic, even as technological progress delivered vaccines in an extraordinarily compressed timeframe. As bluntly stated by the aforementioned author and futurist William Gibson, "The future is already here – it's just not evenly distributed." The prospect of AI has been regarded by humans as the ultimate in engineered advancement, but pitfall-laden owing in no small part to science fiction-based antecedents and echoed by warnings of many scientists, engineers and big thinkers alike. In regard to cosmic threats, we have the technology to deflect potentially impacting asteroids, but lack follow through and funding from the higher-ups to implement. And of course, the sole technological species of our own planet struggles to find ways of shifting to clean, renewable energy to support a sustainable climate as a response to the Anthropocene transition, a responsibility involving every individual.

History has shown that intraspecies competition and more importantly, collaboration, has led us towards the highest peaks of invention. And yet, we prolong notions that seem to be the antithesis of long-term sustainable growth. Racism, genocide, inequity, sabotage… the list sprawls. Some of these have the human condition to blame as a causation - perhaps the subliminal urge of conquest, of imperialism, of victory over an "evil enemy" at all costs. It's clear generations of world leaders have fallen prey to these desires. We can attribute beneficial events to these urges as well: the diffusion of our ancestors across continents and oceans, globalization and interconnectivity, etc. However, humans have surpassed a certain level of civilization that would excuse the Freudian *id* to gratify harmful appetites. Competition for resources exists only as long as hoarding, perceived scarcity and sciolism exist. We certainly have the means to work towards a robust and permanent society. Efficient agriculture, infrastructure, technological improvements and candid leadership are a start in the right direction, towards the diminishment and eventual erasure of our more halting habits. We must consider further measures, especially in these precarious times. It begins with collaboration.

In the perspective of global behavior, the discovery of a planet-rich Universe has rendered less a question of whether aliens exist, but rather in the occurrence that they (statistically, at least) likely do, and are we in a sufficiently stable position to receive such news. But when it comes to foundational questions, the former does present interesting inquiry. If extraterrestrial intelligence does exist, humanity must self-improve on nearly all accounts to meet and even surpass such others. On the flip side, if intelligent life does not appear and perhaps never was, we have some other more philosophical difficulties to juggle – but no less daunting. Our lives are *not* expendable. We have been treating casualties as casual, nukes as necessary, and large-scale death as inevitable events. Life - human life, the lives of our delicate biomes and the millions of species which inhabit it - is unique and so incalculably precious. Carelessness leading to failure is not an option. We are the only ones who can help ourselves; there should be no expectation of mentors or saviors to step down from the sky on our behalf.

Finally, taking into account impeding concepts like the dismissive "Giggle Factor" and imperialistic desires, it is indeed difficult to stamp out problematic human conditions when handling existential disasters. Education and experience have always valued immediate results over playing the long game, and short-sighted agents have constantly been limiters towards humanity's true potential. Hence, the vantage points most take towards the Great Filter have been undermined with ignorance. To overcome these barriers, both the individual and the institution must bring about awareness, and in turn, reform to higher ideals. Indeed, by striving for far-reaching goals we, as a species, may come untangle ourselves from historical troubles. Setting our sights on becoming a Kardashev Type I civilization, perhaps achievable in little more than the time taken to go from the first practical steam-driven engines to the present, would be a 'giant leap for humankind' in the right direction. Attainment of Type I status would all but assure any Great Filter has been successfully overcome, unfolding an all but unlimited future for humanity. For a Type I civilization, the boundary of our Solar System would perhaps be as easily traversed as present-day shorelines of Earthly continents. There is no known theoretical limit to how far humanity could progress into the distant future, this given the effectively inexhaustible supplies of matter and energy which lay beyond Earth's fragile atmosphere in a Universe with far more time ahead of it than the mere 13.8 billion years which have already dropped into its past. Humanity evolving to a Type II civilization, and even to Type III, is not beyond possibility. To prepare for our journey, we can likely count on the inner Solar System remaining habitable for another few billion years, until the Sun begins to expand towards red giant status. Time enough for humanity to finally make other stars our home.

**Acknowledgement:** This work was conducted at the Jet Propulsion Laboratory, California Institute of Technology, under contract with NASA. We also acknowledge the support from the Jagiellonian University of Poland.